\documentclass[bibauthoryear]{aa}

\usepackage{amssymb}
\usepackage{soul}
\usepackage{graphicx}
\usepackage[varg]{txfonts}
\usepackage{amsmath}
\usepackage{natbib}
\usepackage{xcolor}
\usepackage[colorlinks=true,citecolor=blue,linkcolor=blue,urlcolor=blue,anchorcolor=blue]{hyperref}
\usepackage[normalem]{ulem}

\usepackage{tikz}

\newcommand{\tdelta}[1]{\tilde{\delta}_{#1}}
\newcommand{\Om}[0]{\Omega_m}
\newcommand{\OL}[0]{\Omega_{\Lambda}}
\newcommand{\bigexp}[1]{\exp \left[ #1 \right]}

\begin{document}

	\title{The Most Probable Outer Density Profile from Excursion Set Theory}
	
	\author{Ericka Florio
		\inst{1,2}
		\and
		Vasiliki Pavlidou\inst{2, 3}
		}
	    \institute{Centre for Theoretical Cosmology, Department of Applied Mathematics and Theoretical Physics,
        University of Cambridge, Wilberforce Road, Cambridge CB3 0WA, United Kingdom\\
        \email{eaf49@cam.ac.uk}
	    \and
		University of Crete, Department of Physics \& Institute of
		Theoretical \& Computational Physics, 70013 Herakleio, Greece\\
        \email{pavlidou@physics.uoc.gr}
		\and 
		Institute of Astrophysics,
		Foundation for Research and Technology-Hellas, 71110 Heraklion, Crete, Greece
	}
	
	\date{\today}

	\abstract
	{Measurements of the turnaround radius around galaxy clusters can be used to break the degeneracy between measurements of the present day energy density of matter and dark energy. 
    Korkidis \& Pavlidou showed that the turnaround radius coincides with the first point of deviation between outer density profiles in N-body simulations and the analytic profile predicted by excursion set theory. 
    However, their analytic profile relied on a number of simplifying assumptions, which may each introduce systematic error.}
    {We evaluate the effect of these assumptions on the shape of the analytic profile and its correspondence with simulated outer density profiles. }
    {We relax the key simplifying assumptions and re-derive the mode of the outer density profile from excursion set theory.
    We then numerically resolve the double distribution (DD) across a range of masses and clustering parameters, and compare the numerical mode estimate to the re-derived analytic profile.}
    {We find excellent agreement between our analytic profile and the numerically-realized DD.
    However, our analytic profiles diverge from N-body profiles, and this divergence grows as we relax successive assumptions.}
    {We relate this mismatch to the differing window functions used in the analytic and simulation-based approaches, which respectively yield Markovian and correlated density trajectories.
    We conclude that the analytic profile proposed by Korkidis \& Pavlidou should only be used as a few-parameter effective description of the most-probable outer density profile, with parameters fitted to results of cosmological simulations.}
	

	\authorrunning{Florio \& Pavlidou}
	
	\maketitle
	
	
\section{Introduction} 
\label{sec:introduction}
Precision measurements of Large-Scale Structure (LSS) provide essential constraints on cosmological parameters which are complementary to those provided by the cosmic microwave background. 
For instance, \cite{Euclid2020} describes how the upcoming Euclid mission will provide tight constraints on the parameters of the $\Lambda$CDM model by combining multiple LSS observables, such as galaxy clustering and weak lensing, in one experiment.
However, the $\Lambda$CDM parameters measured by traditional cosmological probes are degenerate with one another, which increases their susceptibility to systematic error.
The development of cosmological probes which can break the degeneracy between $\Lambda$CDM parameters is therefore a key scientific target for the precision cosmology community.

One such probe can be found at the outermost edges of the largest gravitationally-bound structures.
The ``turnaround radius'' ($R_{ta}$) is defined as the radial distance at which the recessional velocity induced by cosmic expansion is exactly canceled by the peculiar infall velocity induced by the halo's gravity.
\cite{PT2014} introduce the turnaround radius as an observational test of the bounded nature of structure growth; \cite{TPT2016} then extend this result to show how the turnaround radius can be used to independently probe the value of $\Lambda$ using local structures alone.
The turnaround radius has been calculated across a range of cosmological models, including in modified and alternative theories of gravity (\cite{F2016, BDRST2017, LVAS2018}) and for non-spherical structures (\cite{GF2019}).

In order to measure the turnaround radius in observational data, we must first develop a rigorous theoretical understanding of the environment where the turnaround radius resides, mainly the outermost region of collapsed galaxy clusters and super-clusters.
The most widely used analytic model of formation between cosmological and galactic scales is excursion set theory, which treats the emergence of structure, parameterized by the local fractional overdensity $\delta$, as a stochastic process which evolves according to a classical random walk. 
\cite{PS1974} first used this framework to predict the mass function of collapsed structures; 
\cite{BCEK1991} then placed the work of Press and Schechter on more solid theoretical ground.

However, the mass function alone is insufficient to estimate the turnaround radius, as it only counts the number of collapsed structures of a given \textit{mass}. 
The turnaround radius lies by definition outside the collapsed part of a structure. To trace it, therefore, the \textit{density distribution} of the environment around the collapsed structure must also be quantified. 
\cite{PF2005} address these issues by deriving from the excursion set formalism
a joint probability distribution function (PDF) of collapsed mass and overdensity on some larger scale, called the "double distribution" or DD. 
The DD gives the number of structures per unit mass and extrapolated overdensity at a given scale factor, and reproduces the Press-Schechter mass function when marginalized over overdensity.
When plotted as a function of overdensity for a fixed mass of the central collapsed object, the DD exhibits a well-defined single peak. 
This observation provides a route from the statistical description of the density field to a prediction for the density profile surrounding collapsed structures of a given mass: the calculation of a "most probable outer density profile" encoded by the mode of such slices of the DD. 

\citet{KP2024} followed this route by deriving the mode under simplifying assumptions, thereby obtaining a most probable profile that is independent of central collapsed mass: the universal-scaling, or US profile. 
Physically, this profile is valid only for scales where shell-crossing has never occurred, i.e. outside the turnaround radius. 
\citet{KP2025} have shown that the radius at which the most probable profile extracted from full N-body cosmological simulations deviates from the DD-derived most-probable profile indeed coincides with the physical turnaround radius.
(see e.g. Fig. \eqref{asmp:struct-in-struct} of \citealp{KP2025}).
The simplifying assumptions adopted by \cite{KP2024} in their derivation of the US profile are the following. 
\begin{enumerate}
    \item \label{asmp:order-of-operations} \textbf{Order of operations:} The mode of the distribution was found by first differentiating the DD with respect to linearized overdensity, then converting the linearized overdensity to non-linear overdensity $\delta_l$ via the spherical collapse model. However, the mathematically correct operation would have been to first transform the DD to a distribution with respect to non-linear overdensities, then take the mode to find the most probable profile. The reverse order of operations greatly simplifies the algebra, but is expected to produce some inaccuracy; 
    \cite{KP2024} argued that the close agreement between the resulting profile and N-body simulations suggested that this inaccuracy was small.    
    \item \label{asmp:struct-in-struct} \textbf{The ``structure-in-structure'' term:} \cite{KP2024}  dropped a term in the DD which corrects for halos that are included in still larger gravitationally-bound structures. 
    They argued that the contribution of this structure-in-structure term is minimal in their context, as they focused on the largest possible structures.
    \item \label{asmp:spher-collapse-approx} \textbf{Spherical collapse approximation:} An approximation to the spherical collapse model, given by Eq. (C7) of \cite{PF2005}, was used to convert between linearized and non-linear over-densities. \citet{KP2024} showed that the approximate conversion should produce an error of a few percent in the final result.
    \item \label{asmp:power-law-approx} \textbf{Power-law approximation:} The mass variance of the density field was approximated as a power law, where the slope was determined by a fit to the mass range of the particular sample of simulated clusters used in the comparison with N-body simulations. 
    \item \label{asmp:mass-dependent-us-profile} \textbf{Sub-leading mass dependence:} Partway through the derivation of the most probable profile, 
    \cite{KP2024} 
    assume that 
    \begin{align*}
        \frac{\tdelta{0,c}^2}{S(m) - S(\beta m)} \gg 1,
    \end{align*}
 
    such that they can drop a term which would otherwise impart mass dependence, which was not seen in simulations, on the most probable profile (see \S \ref{subsec:mass-dependent-us-mode} for more details).
    This approximation should produce a maximum error of $\sim 17\%$ within the mass range of simulated clusters these authors compared their results against.
\end{enumerate}
These assumptions were introduced to obtain a tractable, approximately mass-independent profile, motivated by the near-universality of simulated outer profiles for cluster-scale halos. The analytic US profile derived by \cite{KP2024} was found to indeed provide a good approximate fit to simulated outer density profiles beyond the turnaround radius. 

The central question of this work is whether this agreement with simulations survives when the simplifying assumptions used to derive the US profile are relaxed. Counterintuitively, we find that it does not: the corrected prediction agrees with the profile derived numerically directly from the  double distribution, but departs more strongly from the simulated profiles.
We conclude that, by applying the proper transformation to the double distribution, we have amplified a source of error within the framework of excursion set theory itself, which we argue is linked to the Markovian assumption underlying the excursion-set treatment of structure formation on the largest scales. 
We present evidence from the literature to support this conclusion, and provide an outline for future work which could be used to both establish our conclusion more concretely, and to derive the outer density profile from a more robust version of excursion set theory. We finally establish that the analytic profile formula proposed by \citet{KP2024} works well as and only as a {\em few-parameter effective description of outer density profile statistics}, and we encourage its use as such, with parameters fitted to simulated data.

This paper is organized as follows. 
In \S \ref{sec:formalism} we present our semi-analytic results for the mode of the double distribution. 
In \S \ref{sec:results} we provide validation results for our semi-analytic most-probable profile.
We demonstrate the departure of the double distribution from the US profile, and rule out assumptions \eqref{asmp:struct-in-struct}--\eqref{asmp:power-law-approx} in producing this discrepancy.
In \S \ref{sec:discussion}, we argue that this discrepancy arises from the choice of window function used in excursion set theory and in the extraction of profiles from simulations, and present evidence from the literature to support this conclusion.

\section{Formalism} 
\label{sec:formalism}
\subsection{Outline of the double distribution derivation}
\label{subsec:dd-derivation}

\cite{KP2024} derive a most probable profile from the double distribution which displays universality in mass. 
The double distribution is derived by first introducing into excursion set theory a clustering parameter $\beta>1$ associated with a collapsed mass $m$, such that the mass surrounding the collapsed object at a certain radius is given by $\beta m$. In essence, $\beta$ describes the additional mass residing in the collapsed object's environment.
In standard excursion set theory, the overdensity $\tdelta{l}$ follows a classical random walk trajectory through matter variance space, parameterized by $S(m)$.\footnote{$S(m)$ describes the variance of the matter field as a function of mass, and is related to the measurable $\sigma_8$ parameter by $\sigma_8 = \sqrt{S(m_8)}$.}
Eventually this overdensity reaches an ``absorbing barrier'' $\tdelta{c}$, the overdensity of a collapsing structure at the time of collapse. 
After this point, the overdensity is considered part of the collapsed structure and is removed from consideration.
To derive the double distribution, \cite{PF2005} instead consider a random walk through the ``environment space'' parameterized by $S(\beta m)$.
The variance $S(\beta m)$ is incremented in the standard way, by applying a sharp k-space window function to the stochastic density field at successive values of $S(\beta m)$. 

The double distribution is given in Eq. (25) of \cite{PF2005} as
\begin{align}
    \label{eq:double-distribution}
    \nonumber\frac{dn}{dmd\tdelta{l}}(m, \tdelta{l}, \beta, a) = \frac{\rho_{m,0}}{m}\left[\frac{\tdelta{0, c}(a) - \tdelta{l}}{[S(m) - S(\beta m)]^{3/2}}\right] \left|\frac{dS}{dm}\right|_m \\\nonumber  \\
    \nonumber \times \bigexp{-\frac{(\tdelta{0,c}(a) - \tdelta{l})^2}{2[S(m) - S(\beta m)]}}\\ \nonumber \\
    \times \frac{\bigexp{-\frac{\tdelta{l}^2}{2S(\beta m)}} - \bigexp{-\frac{(\tdelta{l} - 2\tdelta{0,c}(a))^2}{2S(\beta m)}}}{2\pi \sqrt{S(\beta m)}}.
\end{align}
Here, $\tdelta{c,0}$ is related to the collapse overdensity $\tdelta{c}$ through the linear growth factor:
\begin{align*}
    \tdelta{c,0} = \tdelta{c}(a)\frac{D(a_0)}{D(a)}
\end{align*}
where 
\begin{align*}
    D(a) = &\frac{\sqrt{a(1-\Om - \OL)+ \Om + \OL a^3}}{a^{3/2}}\\ 
    &\times \int_0^a \left(\frac{x}{x(1-\Om - \OL) + \Om + \OL x^3}\right)^{3/2} dx.
\end{align*}
The collapse overdensity is universal given a particular underlying cosmology, specified by $(\Om, \OL)$.

The double distribution can be broken into two parts, which each arise from different aspects of the excursion set theory approach. 
The first part represents the fraction of points which can be found between an overdensity $\tdelta{l}$ and $\tdelta{l} + d\tdelta{l}$ on a smoothing scale $S(\beta m)$, assuming the trajectory of $\tdelta{l}$ begins at 0. 
This term is given by
\begin{align}
    \label{eq:random-walk-pdf}
    \frac{\bigexp{-\frac{\tdelta{l}^2}{2S(\beta m)}} - \bigexp{-\frac{(\tdelta{l} - 2\tdelta{0,c}(a))^2}{2S(\beta m)}}}{\sqrt{2\pi} \sqrt{S(\beta m)}}.
\end{align}
The first term of this expression encapsulates the Gaussian nature of the linearized overdensity field.
The second term is the ``structure-in-structure'' term, and corrects for the possibility that the given overdensity has been encompassed by a larger collapsing structure.

The second component of the double distribution describes the conditional probability that a particular overdensity $\tdelta{l}$ has already been absorbed by the absorbing boundary.
This term is found by performing the integral of Eq.\eqref{eq:random-walk-pdf} from $-\infty$ to $\tdelta{c,0}$, which gives the number of trajectories yet to cross the boundary, and subtracting this from 1.
The derivative with respect to $m$ is taken, giving the conditional probability 
\begin{align}
    \label{eq:double-distribution-conditional-component}
    \nonumber \frac{dP(>m, a)}{dm} = \frac{\tdelta{0, c}(a) - \tdelta{l}}{\sqrt{2\pi}[S(m) - S(\beta m)]^{3/2}}& \left|\frac{dS}{dm}\right|_m\\
    &\!\!\!\!\!\!\!\!\times \bigexp{-\frac{(\tdelta{0,c}(a) - \tdelta{l})^2}{2[S(m) - S(\beta m)]}},
\end{align}
which is also given by Eq. (14) of \cite{PF2005}.
We return to this part of the double distribution derivation in \S \ref{sec:discussion} to evaluate its impact on our final comparison to simulations.

By applying assumptions \eqref{asmp:order-of-operations}--\eqref{asmp:mass-dependent-us-profile} to Eq.~\eqref{eq:double-distribution}, \cite{PF2005} derive the following outer density profile which is given by the mode of this distribution.
This profile, which we will refer to as the universal-scaling or US profile, is given by
\begin{align}
    \label{eq:universal-scaling-profile}
    \hat{\rho}_{us} \equiv \frac{\rho_{avg}}{\rho_m} = (1-\beta^{-\gamma})^{-\tdelta{c}},
\end{align}
where $\hat{\rho} \equiv \rho/\rho_m$ and $\gamma$ is the power-law scaling of $S(m)$ in its power law approximation (see Eq.~\eqref{eq:S-power-law} below). The implied mass-independence was desirable because the analysis of N-body simulations had indicated that the outer density profiles of the high-mass halos of galaxy clusters are approximately universal. The US profile was therefore derived under assumptions chosen to recover this observed universality. 
In the following sections, we will describe the mathematics used to relax assumptions \eqref{asmp:order-of-operations} and \eqref{asmp:spher-collapse-approx}--\eqref{asmp:mass-dependent-us-profile}.

\subsection{Mode of the transformed distribution}
\label{subsec:mode-of-transformed-distribution}

In order to address assumption \eqref{asmp:order-of-operations}, we transform the double distribution from a function of the linearly-extrapolated overdensity $\tdelta{l}$ to a function of the non-linear matter density $\hat{\rho}$.
This transformation is given by
\begin{align}
    \label{eq:transformed-double-distribution}
    \frac{dn}{d\hat{\rho} dm} = \frac{dn}{d\tdelta{l} dm}\left(\frac{d\hat{\rho}}{d\tdelta{l}}\right)^{-1},
\end{align}
where the derivative factor is derived from the particular choice of conversion between linear and non-linear overdensity used.
In Apdx.~\ref{apdx:transformed-mode-derivation}, we derive the following polynomial representation of the mode of the properly-transformed double distribution:
\begin{align}
    \nonumber 2A'\tdelta{c} X^3 + (2A'\eta - 2B'\tdelta{c})X^2 - (2B'\eta + &2\tdelta{c} + \tdelta{c}^2)X \\
    \label{eq:transformed-mode-polynomial}
    &- \eta(1+\tdelta{c}) = 0,
\end{align}
where $A'$ and $B'$ are constants dependent on $S(m)$ and $S(\beta m)$; $\eta \equiv \tdelta{0,c} - \tdelta{c}$; and $X\equiv \hat{\rho}_{tf}^{-1/\tdelta{c}}$, where we denote the mode of the properly-transformed double distribution as $\hat{\rho}_{tf}$.

\subsection{Full spherical collapse}
\label{subsec:full-spherical-collapse}

In order to match the results from the double distribution to fully-nonlinear N-body simulations, one must carefully choose how one converts linearly-extrapolated over-densities $\tdelta{l}$ to non-linear over-densities $\delta_l$.
The full conversion arises from the spherical collapse model; \cite{KP2024} employ an approximation given by
\begin{align}
    \label{eq:approx-spherical-collapse}
    \tdelta{l}\approx \tdelta{c}\left(1-(1+\delta_l)^{-1/\tdelta{c}}\right) = \tdelta{c}(1-\hat{\rho}^{-1/\tdelta{c}}),
\end{align}
where we have used the non-linear relation $\delta_l = \hat{\rho} - 1$.

In order to determine the impact of this approximation on the final profile $\hat{\rho}$, we implemented the spherical collapse conversion and compared this to the approximate conversion, Eq.~\eqref{eq:approx-spherical-collapse}. 
This non-linear conversion can be determined from equations presented in Apdx. B and C of \cite{PF2005}; we will briefly outline the conversion here.
We will compute the conversion $\hat{\rho} \rightarrow \tdelta{l}$, as this is required to resolve the untransformed distribution Eq.~\eqref{eq:double-distribution} and to determine the derivative term in Equation \eqref{eq:transformed-double-distribution}.
We would eventually like to use P\&F Eq.~(C28),
\begin{align}
    \label{eq:kappa-to-tdelta}
    \tdelta{l} = \kappa \frac{3A[(2\omega)^{1/3}a]}{(2\omega)^{1/3}},
\end{align}
to find $\tdelta{l}$ from $\kappa$, the extrinsic curvature associated with a particular overdensity which is a function of $\hat{\rho}$.
Here $\omega \equiv \OL/\Om$ depends only on the chosen cosmology.

We begin by noting that P\&F Eq.~(B2), while first presented in the context of Einstein-de-Sitter cosmology, is additionally applicable to the $\Lambda$CDM cosmology within which we work.
This equation converts the non-linear matter density $\hat{\rho}$ into $a_p$, the local scale factor corresponding to the perturbation, which obeys a Friedman equation with non-zero curvature $\kappa$.
This relation is given by
\begin{align}
    \label{eq:rho-to-ap}
    a_p = \frac{a}{\hat{\rho}^{1/3}}.
\end{align}
We must convert $a_p \rightarrow a_{p,ta}$, the perturbation scale factor at the point of turnaround.
We can then re-arrange the polynomial equation P\&F Eq.~(C3), which is derived from the ratio of the background and perturbative Friedman equations. 
This produces
\begin{align}
    \label{eq:apta-to-kappa}
    \kappa = \frac{\omega a_{p,ta}^3 + 1}{a_{p,ta}}.
\end{align}

The conversion between $a_p \rightarrow a_{p,ta}$ is determined from the full solution to the Friedman equation for $a_p$ in terms of $a$, given by P\&F Eq.~(C5). 
We will work with the integral form of this solution; as we are most interested in studying profiles of structures measured near $z=0$, after $a_{ta}$ has passed, we will use the second branch of P\&F Eq.~(C15), which is
\begin{align}
    \label{eq:ap-to-apta}
    \omega^{-1/3}\sinh(2{\cal V}_1(1,\mu) - {\cal V}_1(r,\mu))^{2/3} - a = 0,
\end{align}
where $\mu \equiv (\omega a_{p,ta}^3)^{-1}$, $r \equiv a_p/a_{p,ta}$ and ${\cal V}_1$ is the first vacuum integral, given by
\begin{align}
    \label{eq:first-vacuum-integral}
    {\cal V}_1(r, \mu) = \frac{3}{2} \int_0^r \frac{\sqrt{x}\ dx}{\sqrt{(1-x)(-x^2-x+\mu)}}.
\end{align}
We evaluate this integral numerically, then solve Eq.~\eqref{eq:ap-to-apta} iteratively for $a_{p,ta}(a_p)$ using Newton's method until a specified level of convergence is achieved.

We additionally note that ${\cal V}_1$ is only defined where $0 \leq r \leq 1$ and $\mu \geq 2$. The bound on $\mu$ can be used to put bounds on $\hat{\rho}$ for which this conversion can be used.
First, note that by the definition of $a_{p,ta}$, $0 \leq a_p/a_{p,ta} \leq 1$ which implies $0 \leq a_p \leq a_{p,ta}$. 
Now, $\mu \geq 2 \Longrightarrow a_{p,ta} \leq (2\omega)^{-1/3}$ from the definition of $\mu$.
Thus $a_p \leq (2\omega)^{-1/3}$, and from Eq.~\eqref{eq:rho-to-ap} this implies
\begin{align}
    \label{eq:rho-hat-bounds}
    \hat{\rho} \geq 2\omega.
\end{align}
In \S \ref{sec:results}, we compare this bound on $\hat{\rho}$ to the effective bound due to the stiffness of Eq.~\eqref{eq:ap-to-apta} near $2\omega$.

\subsection{Matter variance from Bardeen's transfer function}
\label{subsec:bardeens-transfer-function}

In order to evaluate both the transformed and universal-scaling profiles, one must choose a functional form for $S(m)$. \cite{KP2024} choose a power-law approximation
\begin{equation}
    \label{eq:S-power-law}
    S_{pla}(m) = S_0 \left(\frac{m}{m_0}\right)^{-\gamma},
\end{equation}
where $S_0$ sets the scale of $S_{pla}$ and $m_0$ sets the scale for $m$. 
Both of these parameters cancel during the derivation of the US profile, leaving only $\gamma$ as a free parameter which must be fit to the simulated cluster groups under study. 
This power law approximation has the convenient property that it does not transfer mass dependence to the US mode.

We implement the power law approximation and additionally find the more precise form of the matter field variance, $S_{B}(m)$. 
This matter variance can be found by evaluating the integral form of $S(m)$, which applies a transfer function to the power spectrum of initial conditions predicted by cosmic inflation, $P_{inf}(k) \propto k^{n}$ where $n$ is the primordial spectral index:
\begin{align}
    \label{eq:full-S}
    S_{B}(m) = \sigma_8^2 \frac{\int_{0}^{k(m)} T^2(k) k^{n+2} dk}{\int_0^{k(m_8)} T^2(k) k^{n+2} dk},
\end{align}
where $T(k)$ is the Bardeen transfer function derived in \cite{BBKS1986}.
It reads
\begin{align*}
    T(k) = \frac{\ln(1 + 2.34 q)}{2.34 q} [1 + 3.89q + (16.1 q)^2 + (5.46 q)^3 \\ + (6.71 q)^4]^{-1/4},
\end{align*}
where $q \equiv k/\Om h^2 Mpc^{-1}$. 
The form of $k(m)$ is derived from the integral of the configuration-space representation of the sharp k-space window function, and reads
\begin{align}
    k(m) = \left( \frac{6 \pi^2 \rho_{m,0}}{m} \right)^{1/3}
\end{align}
We evaluate Eq.~\eqref{eq:full-S} numerically for each value of mass in our domain.

\subsection{Adding mass dependence to the US profile}
\label{subsec:mass-dependent-us-mode}

After finding the derivative of Eq.~\eqref{eq:double-distribution} with respect to $\tdelta{l}$, \cite{KP2024} derive an expression for the most probable density profile in terms of a polynomial of $\tdelta{l}$:
\begin{align}
    \label{eq:tdelta-l-quadratic}
    \nonumber \tdelta{l}^2 \left( \frac{1}{S(m) - S(\beta m)} + \frac{1}{S(\beta m)}\right) - \tdelta{l}\tdelta{c,0}\left( \frac{2}{S(m) - S(\beta m)} + \frac{1}{S(\beta m)} \right) \\+ \frac{\tdelta{c,0}^2}{S(m)-S(\beta m)} - 1 = 0.
\end{align}
The authors then argue that, as $\frac{\tdelta{c,0}^2}{S(m)-S(\beta m)} \gg 1$ for the region of parameter space they consider, they can neglect the $(-1)$ term at the end.
This is convenient, as later algebraic steps would involve multiplying both sides of this equation by $S(m)$ and $S(\beta m)$, which would introduce a mass dependent term in the place where this $(-1)$ had been.
We relax this assumption by solving for the roots of Eq.~\eqref{eq:tdelta-l-quadratic} directly. In \S \ref{sec:results}, we compare this mass-dependent mode with the universal-scaling mode.

\section{Results} 
\label{sec:results}
In this section, we present results obtained using a python repository we have developed specifically to study the full double distribution and its properties.\footnote{see Acknowledgments for a GitHub link.}
In this repository, we numerically resolve the double distribution on the relevant section of $(\hat{\rho}, m, \beta)$ parameter space.
We then implement the mathematical formulae described in \S \ref{sec:formalism} in order to successively relax the assumptions listed in \S \ref{sec:introduction}. 
This computational pipeline serves two purposes: to validate the semi-analytic calculation against the direct numerical evaluation of the DD, and to determine whether relaxing the simplifying assumptions improves agreement with simulated profiles.

All of the results involving matter density are plotted with respect to $\hat{\rho}$, rather than $\tilde{\delta}_{l}$.
We label all axes representing the double distribution, either Eq.~\eqref{eq:double-distribution} or \eqref{eq:transformed-double-distribution}, as $P_n$, the probability distribution of the number density of structures.
We use the following cosmological parameters: $\Om = 0.27$, $\OL = 1 - \Om$, $\sigma_8 = 0.84$, $m_8 = 2\times 10^{14}M_{\sun}$, $h = 71 $ km/s/Mpc, $\rho_c = 2.78\times 10^{11} h^2\ M_{\sun}/\text{Mpc}^3$ and $n=1$. 
Under this cosmology, $\tdelta{c} = 1.6758$.
All figures except for Fig.~\ref{fig:non-linear-conversion} use 100 values of $\hat{\rho}$ in the range $(0.001, 10)$.
We choose values of $\beta$ and $m$ to match the section of parameter space studied in \cite{KP2024}.

\begin{figure}
    \centering
    \includegraphics[width=1.0\linewidth]{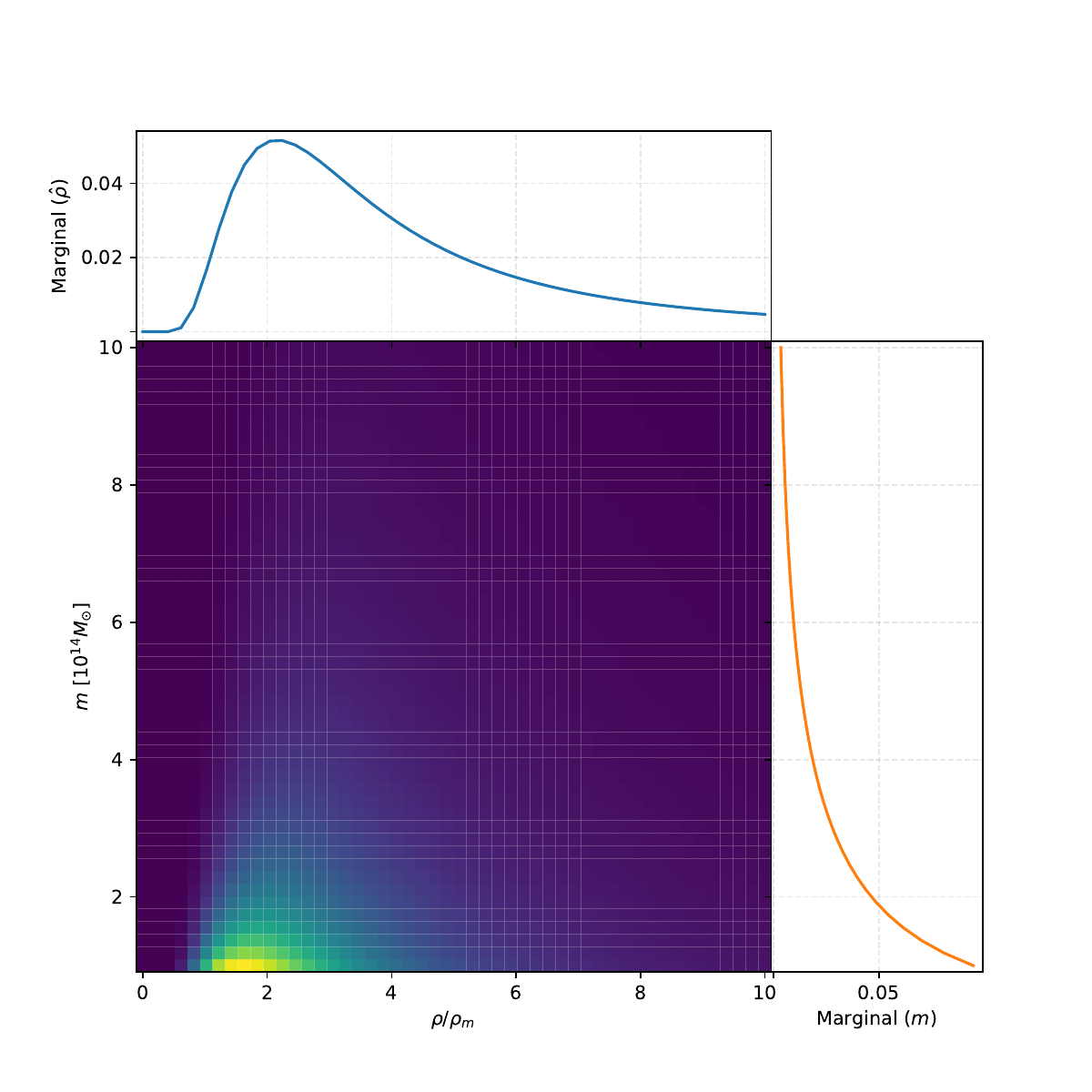}
    \caption{A heatmap of the joint double-distribution, Eq.~\eqref{eq:double-distribution}, with respect to mass (scaled by $10^{14} M_{\sun}$) and rescaled matter density $\hat{\rho}$. Marginal distributions are shown in inlaid boxes.}
    \label{fig:joint-pdf}
\end{figure}

Fig.~\ref{fig:joint-pdf} shows the two-dimensional heatmap of the transformed double distribution described by Eq.~\eqref{eq:transformed-double-distribution}, along with the marginal distributions along each axis. 
This distribution is normalized on the two-dimensional parameter space shown, such that its integral over this parameter grid is equal to 1.
This distribution has a peak at low values of mass, then decays into a broad distribution at high mass, which agrees qualitatively with the results shown in \cite{KP2024}, Fig. 1.

\begin{figure}
    \centering
    \includegraphics[width=1.0\linewidth]{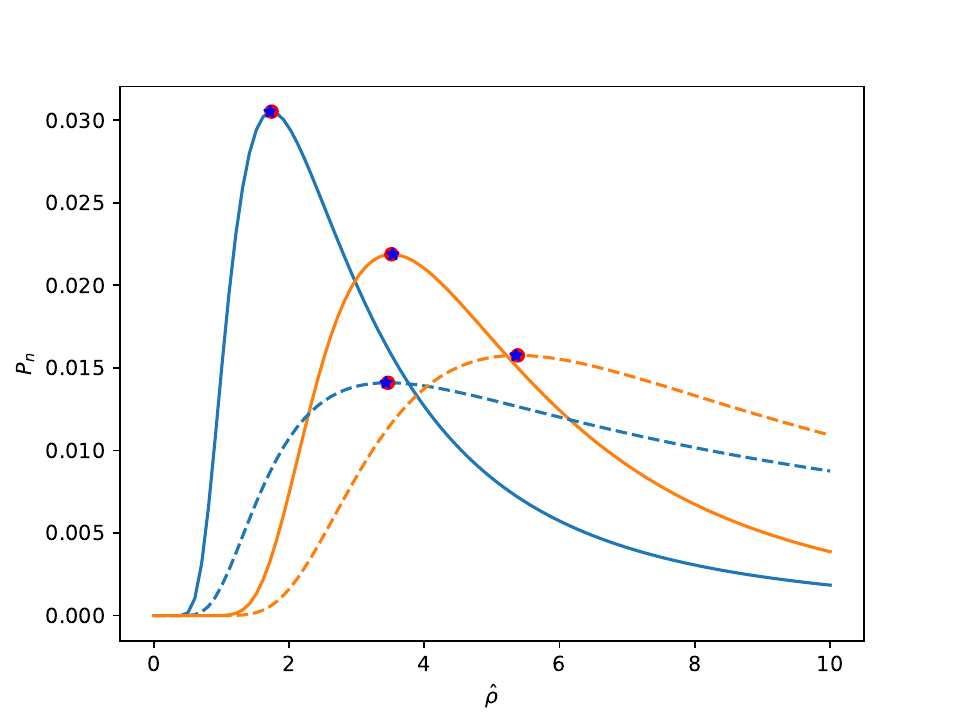}
    \caption{Conditional slices of the joint distribution at $m=1.3\times10^{14} M_{\sun}$ (blue) and $1.7\times 10^{15} M_{\sun}$ (orange). Dashed lines show the untransformed double distribution, and solid lines show the transformed double distribution. Red dots indicate the analytic mode of each curve, and blue stars indicate the numeric mode of each curve.}
    \label{fig:joint-pdf-slice}
\end{figure}

\begin{figure*}
    \begin{tikzpicture}
        \node at (0,0) {\includegraphics[width=0.49\linewidth]{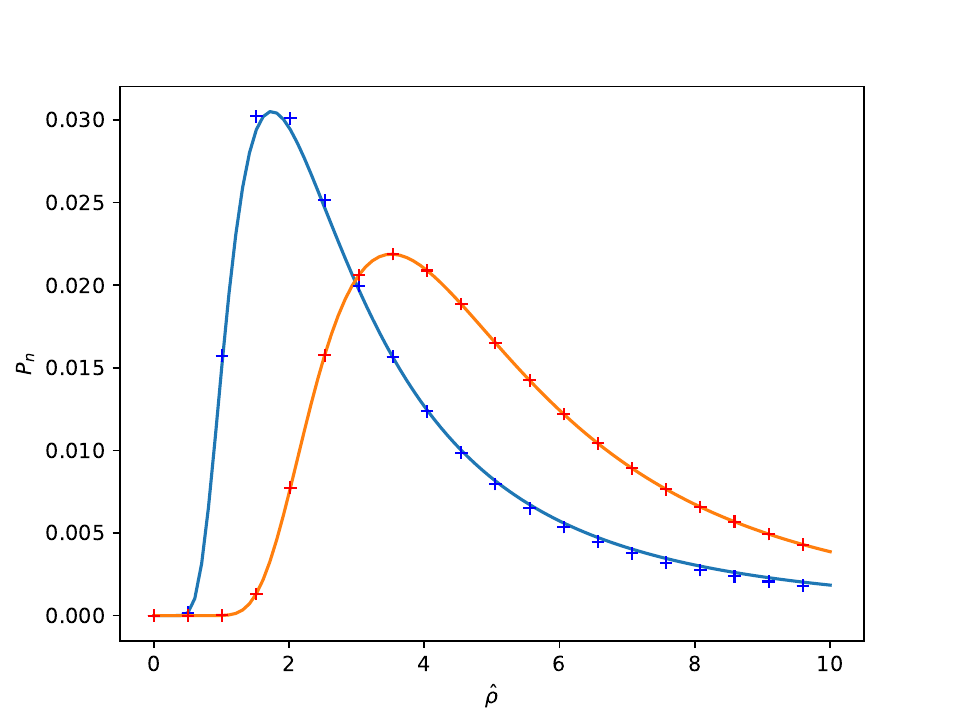}};
        \node at (2.5, 2.0) {(a)};
    \end{tikzpicture}
    \hfill
    \begin{tikzpicture}
        \node at (0,0) {\includegraphics[width=0.49\linewidth]{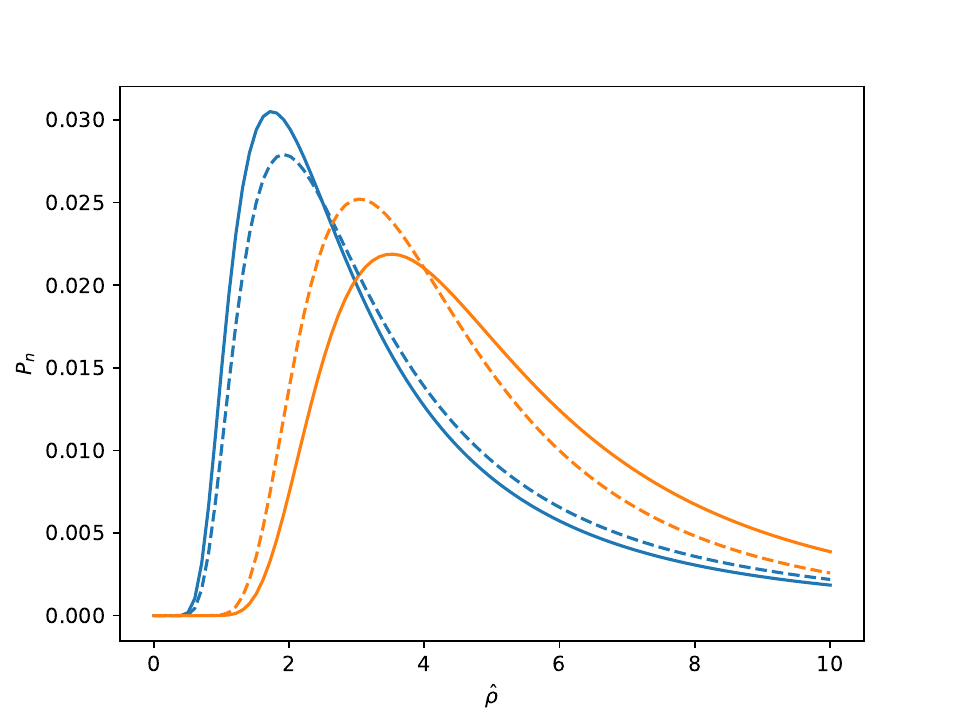}};
        \node at (2.5, 2.0) {(b)};
    \end{tikzpicture}
    \caption{The transformed double distribution, sliced at $m=1.3\times 10^{14} M_{\sun}$ (blue) and $m=1.7\times 10^{15} M_{\sun}$ (orange/red). The solid lines show the transformed distribution without the structure-in-structure term and with a power law approximation assumed. The crosses in panel (a) show the transformed double distribution with the structure-in-structure term included. The dashed lines in panel (b) show the double distribution where the Bardeen transfer function, Eq.~\eqref{eq:full-S}, has been used to calculate $S(m)$.}
    \label{fig:sis-and-pla-comparisons}
\end{figure*}

Figs.~\ref{fig:joint-pdf-slice} and \ref{fig:sis-and-pla-comparisons} show conditional probability distributions, calculated by normalizing the joint double distribution at two values of mass, $1.3\times 10^{14} M_{\sun}$ and $1.7\times 10^{15} M_{\sun}$.
These figures both assume a power-law form for $S(m)$, ignore the structure-in-structure term and use the approximate conversion Eq.~\eqref{eq:approx-spherical-collapse} to convert $\hat{\rho} \rightarrow \tdelta{l}$.
In Fig.~\ref{fig:joint-pdf-slice}, we plot the untransformed double distribution given in Eq.~\eqref{eq:double-distribution} alongside the distribution transformed according to Eq.~\eqref{eq:transformed-double-distribution}.  For both distributions, the red dot marks the analytic mode: it is obtained from Eq.~\eqref{eq:tdelta-l-quadratic} for the untransformed distribution and from Eq.~\eqref{eq:transformed-mode-polynomial} for the transformed distribution. The blue star marks the corresponding numerical mode.
We note that the transformation seems to push the mode of the distribution to lower values of $\hat{\rho}$, which is consistent with the derivative of this conversion shown in Fig.~\ref{fig:non-linear-conversion}.
The numerical and analytical estimates of the modes agree for both the transformed and untransformed distributions.
This provides evidence that our semi-analytic solution is both mathematically correct, and is properly resolved by our python code.

In Fig.~\ref{fig:sis-and-pla-comparisons}, we demonstrate the change in the transformed distribution when both the structure-in-structure term is included in the PDF, and when the Bardeen transfer function is used to calculate $S(m)$. 
The structure-in-structure term modifies the transformed distribution only slightly, by raising the amplitude of the peak for low values of mass.
However, the overall agreement in the shape of these two distributions suggests that ignoring the structure-in-structure term will minimally impact further results.
Adjusting the form of $S(m)$ impacts the PDF more strongly, primarily by decreasing the separation between modes at different slices of mass.

\begin{figure*}
    \begin{tikzpicture}
        \node at (0,0) {\includegraphics[width=0.49\linewidth]{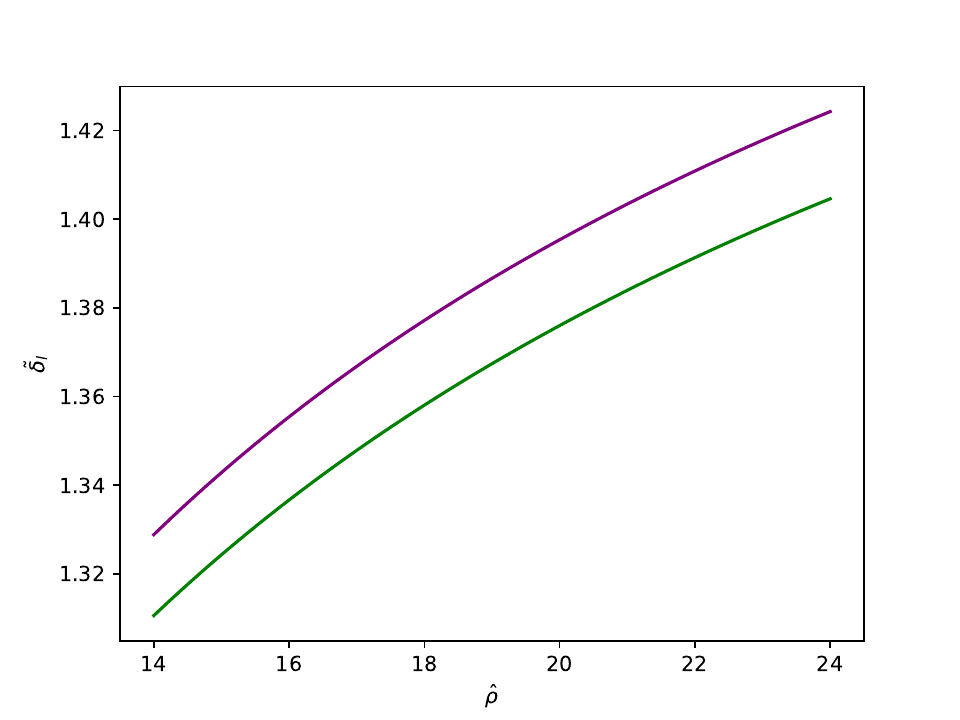}};
        \node at (2.5, -2.0) {(a)};
    \end{tikzpicture}
    \hfill
    \begin{tikzpicture}
        \node at (0,0) {\includegraphics[width=0.49\linewidth]{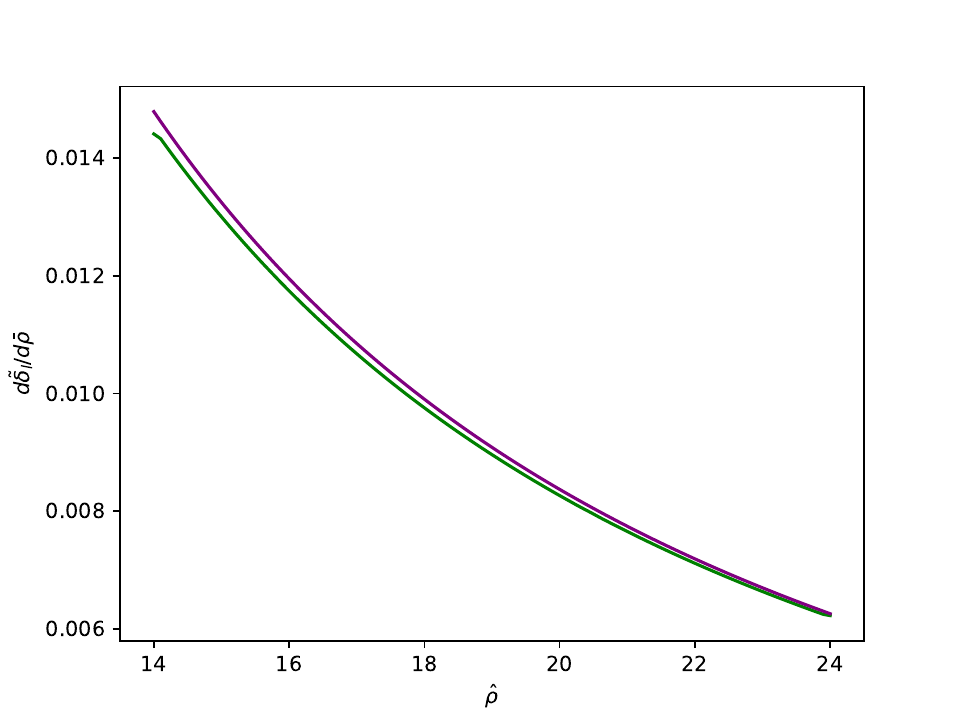}};
        \node at (-2.5, -2.0) {(b)};
    \end{tikzpicture}
    \caption{The transformation $\hat{\rho}\rightarrow \tdelta{l}$, as given by the full solution to the non-linear evolution equation, Eq.~\eqref{eq:kappa-to-tdelta} (green), and the approximate equation Eq.~\eqref{eq:approx-spherical-collapse} (purple). Panel (a) shows the conversion itself, and panel (b) shows the derivative of this conversion.}
    \label{fig:non-linear-conversion}
\end{figure*}

Fig.~\ref{fig:non-linear-conversion} shows a comparison between the approximate spherical collapse conversion, Eq.~\eqref{eq:approx-spherical-collapse}, and the full spherical collapse conversion, Eq.~\eqref{eq:kappa-to-tdelta}.
As shown in \S \ref{subsec:full-spherical-collapse}, $\hat{\rho}$ is bounded from below according to the range over which the first vacuum integral is defined.
We first estimated that $\hat{\rho} \ga 2\omega \approx 5.4$ for this cosmology -- however, due to the stiffness of the solution of Eq.~\eqref{eq:ap-to-apta} near this bound, we were only able to compute solutions within a reasonable number of iterations above $\hat{\rho} \sim 14$.
We were unable to determine how using the full spherical collapse model near the mode of the distribution may affect its location, as the modes of the distributions calculated here all fall well below $\hat{\rho} = 10$.
However we can confirm that, within the range of $\hat{\rho}$ values which can be resolved, the approximate conversion from $\hat{\rho} \rightarrow \tdelta{l}$ produces an error around a few percent across a wide range of $\hat{\rho}$ values, confirming the result in \cite{PF2005}.
Furthermore, the derivative of this approximate conversion matches the true derivative very closely, providing evidence that the inverse derivative used to transform the double distribution is in fact producing a faithful realization of the transformed double distribution if this approximation is used.

\begin{figure}
    \centering
    \begin{tikzpicture}
        \node at (0,0) {\includegraphics[width=1.0\linewidth]{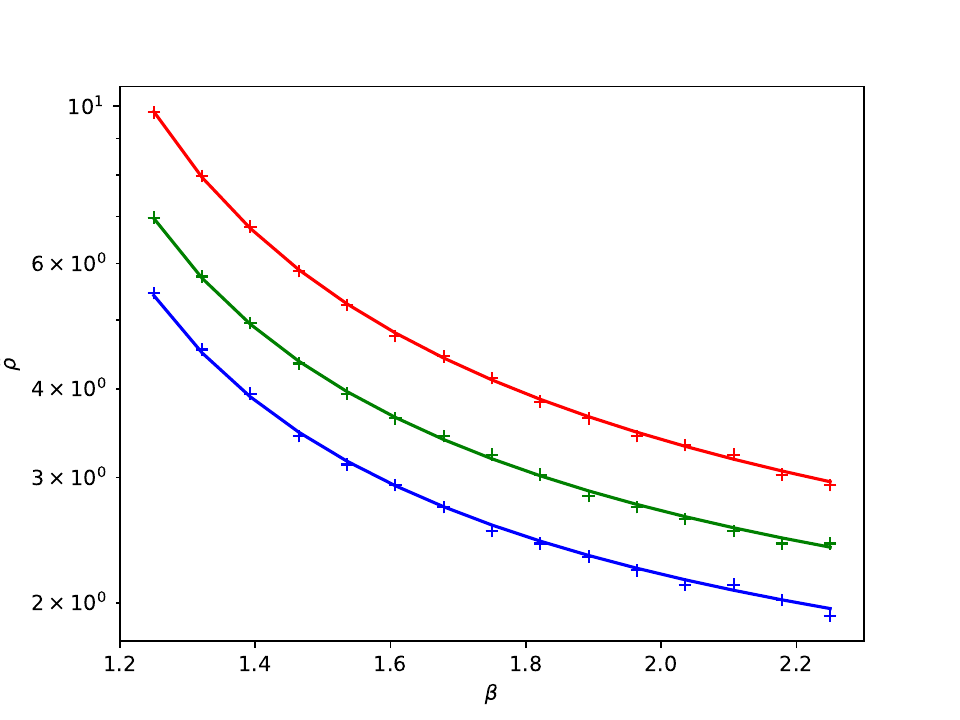}};
        \node[scale=0.25] at (-4.54, -0.03) {<};
    \end{tikzpicture}
    \caption{The most probable density profile for the transformed distribution at $m =$ 1.3 (blue), 5 (green), and 17 (red) times $10^{14} M_{\sun}$. The solid line shows the analytic mode, and the crosses show the numeric mode.}
    \label{fig:profile-analytic-numeric-match}
\end{figure}

\begin{figure}
    \centering
    \includegraphics[trim={50mm 30mm 50mm 30mm}, clip, width=1.0\linewidth]{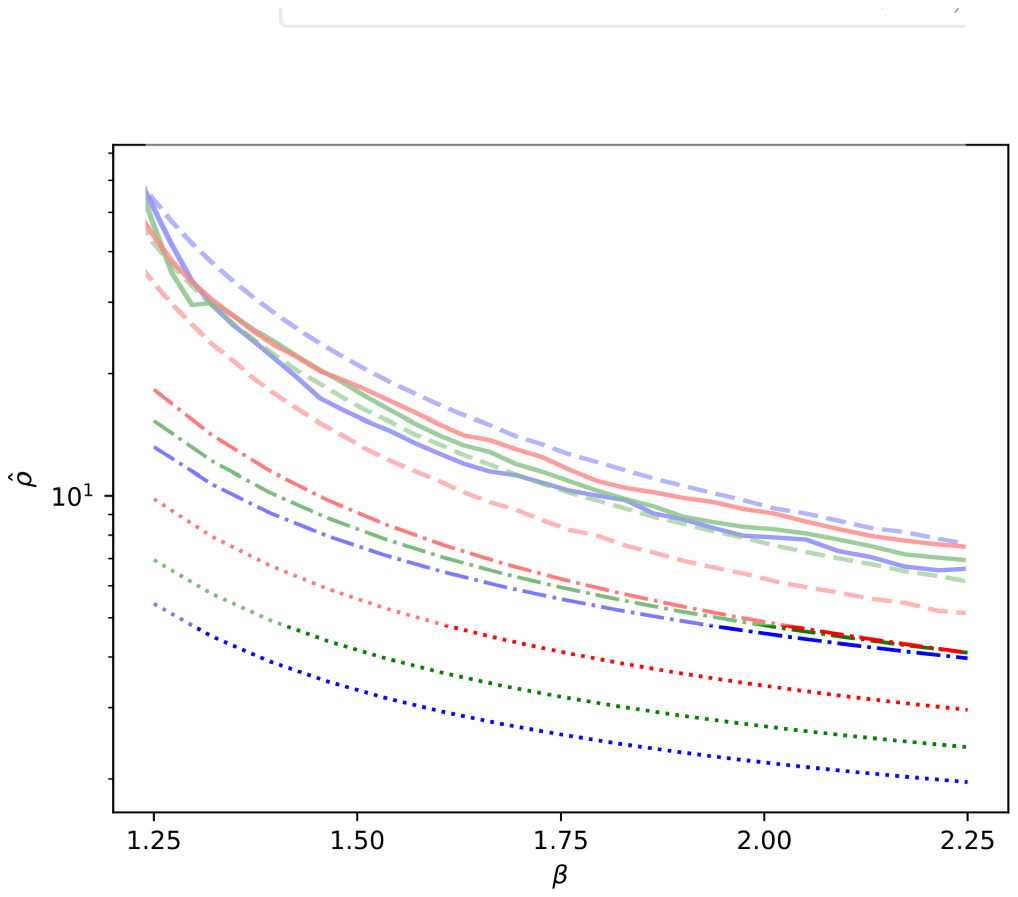}
    \caption{A comparison between both the transformed (dotted) and untransformed (dot-dashed) double distributions' analytic modes, and the results from \cite{KP2024}, showing the profile extracted from N-body simulations (solid) alongside the universal scaling profile for the same range of masses (dashed). Note that all of these modes correspond to masses $m =$ 1.3 (blue), 5 (green), and 17 (red) times $10^{14} M_{\sun}$.}
    \label{fig:profile-comparison}
\end{figure}

Figs.~\ref{fig:profile-analytic-numeric-match} and \ref{fig:profile-comparison} show the most probable profile for $m = (1.3, 5, 17) \times 10^{14} M_{\sun}$. 
These profiles were found by computing the mode of the conditional distribution at the given mass for a range of values of the scaling parameter $\beta$. 
Note that we do not normalize the double distribution over the $\beta$ axis, as $\beta$ is a parameter of this distribution rather than an independent variable.
Both of these figures use 100 values of $\hat{\rho}$ between $(0.001, 10)$ and 15 values of $\beta$ between $(1.25, 2.25)$.
Fig.~\ref{fig:profile-analytic-numeric-match} shows the profile of the transformed double distribution, calculated both numerially and semi-analytically via Eq.~\eqref{eq:transformed-mode-polynomial}.
The agreement between these two curves provides further evidence that Eq.~\eqref{eq:transformed-mode-polynomial} correctly calculates the mode of the transformed double distribution.

Fig.~\ref{fig:profile-comparison} compares four sets of profiles.
The solid and dashed profiles are taken directly from Fig.~5 of \cite{KP2024}; the solid profiles were measured from N-body simulations, and the dashed profiles represent the universal-scaling profile, Eq.~\eqref{eq:universal-scaling-profile}, evaluated using a scaling parameter $\gamma$ fit to each mass group.
The dash-dotted lines represent the analytic mode of the untransformed distribution, Eq.~\ref{eq:tdelta-l-quadratic}.
The dotted lines represent the mode of the transformed distribution, Eq.~\eqref{eq:transformed-mode-polynomial}.
We note that our computation of the US profile did match the results of \cite{KP2024} shown here, but that we have excluded our computation of the US profile from Fig.~\ref{fig:profile-comparison} for the sake of visual clarity.

Although our realizations of the double distribution have passed many internal consistency checks, as described in Figs.~\ref{fig:joint-pdf-slice}--\ref{fig:profile-analytic-numeric-match}, we find an obvious disagreement between the profiles produced directly from the double distribution and the profiles which are measured from N-body simulations.
Furthermore, we find that the profile arising from N-body simulations agrees best with the universal-scaling profile, which was in fact arrived at by applying assumptions \eqref{asmp:order-of-operations}--\eqref{asmp:mass-dependent-us-profile} to the double distribution.
In the following section, we reflect on what this result implies for the applicability of the excursion set theory approach to describing structure formation in simulations.
 
\section{Discussion and Conclusion} \label{sec:discussion}
In this work, we have derived the most probable outer density profile of the largest gravitationally-bound structures directly from the double distribution, which in turn is derived from the excursion set theory framework.
We compared this most probable profile to the results of N-body simulations performed in \cite{KP2024} and to the universal scaling profile, another outer density profile derived from the double distribution which exhibits universality in mass.

We emphasize that the universal scaling profile in fact provides a good fit to simulations, where the parameter $\gamma$ must be fit to the mass range of the structures under study.
Nevertheless, as we removed assumptions from the derivation of the US profile, ostensibly producing a more faithful representation of the result from excursion set theory, we found not only that the profile from excursion set theory diverged from the profile extracted from simulations, but this divergence became larger as we removed more assumptions.
Given that our mode estimation methods have passed multiple internal consistency checks, we now hypothesize that the divergence between theoretical and simulated profiles arises from a deeper assumption within the excursion set framework itself.

Excursion set theory has been questioned by other authors on both theoretical grounds (\cite{ARSC2013}) and for its disagreement with N-body simulations in certain regions of parameter space (\cite{BV1992}).
One particular aspect of excursion set theory that has been highlighted as the source of such issues is the choice of window function.
As explained in \S \ref{subsec:dd-derivation}, excursion set theory uses a sharp k-space window function to increment the random variable $\tdelta{l}$ along the $S(m)$ axis.
This sharp k-space window function has the property that modes which are successively introduced into the evolution of $\tdelta{l}$ are independent of one another.
However, in simulations one must use a top-hat window function in configuration space, which integrates the matter density within spheres of a given radius, to construct profiles, because this window function is the only window function for which it is possible to rigorously define a notion of enclosed mass. 
However, the top-hat window function induces correlations different Fourier modes, and therefore these two procedures should not be expected \textit{a priori} to produce compatible results.

Previous authors have noted this issue, and shown that the choice of window function does indeed impact measures of structure formation.
\cite{BCEK1991} were among the first to point this out, demonstrating a discrepancy in both the qualitative realization of the random density field and in the number of extremely low-and-high-mass objects observed when different window functions are employed.
\cite{LBLP2018} recently proposed a smooth k-space filter, which alleviates to a degree the inconsistent count of low-mass objects produced by the Press-Schechter formalism.
\cite{WSHA2025} confirm that the Press-Schechter inconsistently counts low-mass objects before extending these results to non-spherical objects.
Each of these authors evaluated the impact of these window functions by implementing them in \textit{simulations}; we seek to evaluate whether the sharp k-space window function required by classical excursion set theory would explain our results on \textit{theoretical} grounds alone.

To investigate the origin of this discrepancy, we turn to a specific aspect of the excursion set theory framework laid out in \cite{BCEK1991}. 
They describe how the main result of excursion set theory can be stated as a solution to a stochastic differential equation.
They first show that a random variable $\tdelta{l}$ evolving along $S(m)$ obeys a Fokker-Planck equation.
The solution to the Fokker-Planck equation at a given step returns the PDF from which the next step in the random walk is drawn.
They then use the properties of the sharp k-space window function to set the drift term to zero -- the independence of successive increments makes the walk Markovian. 
Diffusion remains, and the variance grows. 
Therefore, classical excursion set theory actually represents a driftless Markovian process.

We now turn back to the discrepancy observed between the excursion set theory profile and the profile observed in simulations.
We note that an obvious difference appears even for a profile very similar to the universal scaling profile -- that of the solution to Eq.~\eqref{eq:tdelta-l-quadratic}, shown in dot-dashed lines in Fig.~\ref{fig:profile-comparison}.
The only mathematical difference between this profile and the US profile is the inclusion of the constant term $(-1)$ in the quadratic Eq.~\eqref{eq:tdelta-l-quadratic}. 
This constant term arises in the derivation of the double distribution, and specifically in the linear factor $(\tdelta{c,0} - \tdelta{l})$ in Eq.~\eqref{eq:double-distribution-conditional-component}.  
In a classical random walk, the random variable always begins at zero. 
However, we seek to track the trajectories of particles that begin with some value $\tdelta{l} \neq 0$. 
\cite{PF2005} first calculate the classical conditional probability, i.e. the probability that a perturbation has been absorbed by the boundary at time $S(m)$ if it started at $0$, and then perform the substitution $\tdelta{c,0} \rightarrow \tdelta{c,0} - \tdelta{l}$, thereby calculating the odds that a perturbation has been absorbed by the boundary if it started at $\tdelta{l}$. 
\textit{This substitution relies on the Markovian property of the sharp-k walk}: the future evolution depends only on its "current" overdensity. For correlated non-Markovian walks, this need not hold, because the evolution can also depend on preceding steps. 
Before this substitution is made, the linear factor is simply the constant  
$\tdelta{c,0}$, and the term produced in the derivation of 
Eq.~\eqref{eq:tdelta-l-quadratic} would not appear.

These considerations suggest that a derivation of the double distribution from a framework which allows for non-Markovian evolution could establish a link between the error caused by the linear term in Eq.~\eqref{eq:double-distribution-conditional-component} and the error expected in the use of the sharp k-space window function.
Furthermore, such a framework may allow a derivation of the US profile (which fits simulations well and exhibits several appealing theoretical properties) from a more fundamental stochastic theory, without relying on  as many simplifying assumptions.

Previous authors have developed formalisms which incorporate non-Markovian effects into stochastic theories of structure formation.
\cite{MR2010} address this issue by considering a path-integral formulation of excursion set theory, which they then generalize to include sequentially higher-order terms in an expansion about the stationarity of the distribution.
\cite{HP2017} and \cite{H2022} produce Monte-Carlo simulations which resolve the non-Markovian nature of the smoothing process using a top-hat filter in configuration space.
Finally, \cite{LD2020} derive the solution to the Fokker-Planck equation directly.
They recover the Press-Schechter mass function as the stationary solution and then demonstrate that adopting a mass-dependent collapse threshold $\tdelta{c}$ improves the fit to simulations.
We suspect that a re-derivation of the double distribution from the theoretical framework described in \cite{MR2010} or \cite{LD2020} may yield a mathematically rigorous explanation of how the universal scaling profile could produce such a good fit to simulations, when the excursion set theory framework from which it is derived does not.
However such a derivation falls out of the scope of this paper, and we leave such calculation to future work.

\begin{acknowledgements}
We thank Kostas Tassis, Johanna M\"{u}ller and Anna Kyvernitaki-Synani for useful discussions.  
EF acknowledges support by the Gianna Angelopoulos Program for Science Technology and Innovation (GAPSTI). 
This research is funded by the European Union. 
Views and opinions expressed are, however, those of the author(s) only and do not necessarily reflect those of the European Union or the European Research Council Executive Agency. 
Neither the European Union nor the granting authority can be held responsible for them. 
VP's work is supported by an ERC grant, mw-atlas project no. 101166905.	
The code used to produce the results presented here can be found at \url{https://github.com/the-florist/galactic-environment-statistics}.
\end{acknowledgements}

\bibliography{references}

\begin{appendix}
\label{sec:appendices}
\section{Derivation of $\hat{\rho}_{tf}$}
\label{apdx:transformed-mode-derivation}
In this section, we derive an equation for the mode of the double distribution which relaxes assumptions \eqref{asmp:order-of-operations}, \eqref{asmp:power-law-approx} and \eqref{asmp:mass-dependent-us-profile}. Throughout this derivation, we retain assumption (2) and therefore neglect the structure-in-structure term in the double distribution.
We first transform the linearly-extrapolated overdensity $\tdelta{l}$ to the non-linear matter density $\hat{\rho}\equiv \frac{\rho}{\rho_m}$ via the non-linear overdensity $\delta_l$. 
We then find the maximum of the resulting transformed PDF.

First, we note that the double distribution Eq.~\eqref{eq:double-distribution} can be cast in the form 
\begin{align}
    \frac{dn}{dmd\tdelta{l}} = N(\tdelta{0,c} - \tdelta{l})\bigexp{-A\tdelta{l}^2 + 2B\tdelta{l} - C}
\end{align}
where 
\begin{align*}
    N &= \frac{\rho_{m,0}}{2\pi m}\left|\frac{dS}{dm}\right|_m\left(S(\beta m)(S(m)-S(\beta m))^3\right)^{-1/2}\\
    A &= \frac{S(m)}{2S(\beta m)(S(m)-S(\beta m))}\\
    B &= \frac{\tdelta{0,c}}{2(S(m)-S(\beta m))}\\
    C &= \frac{\tdelta{0,c}^2}{2(S(m)-S(\beta m))}
\end{align*}
We will first calculate
\begin{align*}
    \frac{dn}{d\hat{\rho} dm} = \frac{dn}{d\tdelta{l} dm}\frac{d\tdelta{l}}{d\hat{\rho}},
\end{align*}
the transformed double distribution. 
The first term is simply the double distribution, where we will use $\tdelta{l}(\hat{\rho})$ as given by the approximate spherical collapse relation, Eq.~\eqref{eq:approx-spherical-collapse}.
Next, we calculate the derivative of Eq.~\eqref{eq:approx-spherical-collapse}:
\begin{align}
    \label{eq:approx-spher-collapse-derivative}
    \nonumber \frac{d}{d\hat{\rho}}\tdelta{l}(\hat{\rho}) &= \tdelta{c}\frac{d}{d\hat{\rho}}\left[1-\hat{\rho}^{-1/\tdelta{c}}\right]\\
    &= \hat{\rho}^{-1/\tdelta{c}-1}.
\end{align}
Evaluating the double distribution together with this factor gives:
\begin{multline*}
    N\cdot (\tdelta{c,0} - \tdelta{c}(1-\hat{\rho}^{-1/\tdelta{c}}))\hat{\rho}^{-1/\tdelta{c}-1}\\ \bigexp{-A(\tdelta{c}(1-\hat{\rho}^{-1/\tdelta{c}}))^2 + 2B\tdelta{c}(1-\hat{\rho}^{-1/\tdelta{c}}) - C}.
\end{multline*}
To simplify notation, we define $X\equiv \hat{\rho}^{-1/\tdelta{c}}$ and note that $\hat{\rho}^{-1} = X^{\tdelta{c}}$. 
We can then rewrite the transformed PDF in terms of a polynomial multiplied by an exponentiated polynomial:
\begin{align}
    \nonumber &N\cdot (\tdelta{c,0} - \tdelta{c}(1-X))XX^{\tdelta{c}}\\ 
    \nonumber &\ \ \ \ \cdot \bigexp{-A\tdelta{c}^2(1-X)^2 + 2B\tdelta{c}(1-X) - C}\\
    \nonumber &= NX^{\tdelta{c}}\left(\tdelta{c,0} X - \tdelta{c} X + \tdelta{c} X^2 \right)\\
    \nonumber &\ \ \ \ \cdot \bigexp{-A \tdelta{c}^2 + 2A\tdelta{c}^2 X - A\tdelta{c}^2 X^2 + 2B\tdelta{c} - 2B\tdelta{c}X - C}\\
    \nonumber &= NX^{\tdelta{c}}[(\tdelta{0,c}-\tdelta{c})X + \tdelta{c}X^2]\\
    \nonumber &\ \ \ \ \cdot \bigexp{-A\tdelta{c}^2 X^2 + 2\tdelta{c}(A\tdelta{c}-B)X - (A\tdelta{c}^2 - 2B\tdelta{c} + C)}\\
    \label{eq:dd-for-X}
    &= NX^{\tdelta{c}}[\eta X + \tdelta{c} X^2]\bigexp{-A' X^2 + 2B'X - C'},
\end{align}
where we have defined a new set of polynomial coefficients,
\begin{align*}
    \eta &\equiv \tdelta{0,c} - \tdelta{c},\\
    A' &\equiv A\tdelta{c}^2\\
    B' &\equiv \tdelta{c}(A\tdelta{c}-B)\\
    C' &= A\tdelta{c}^2 - 2B\tdelta{c} + C.
\end{align*}

We then take the derivative of Eq.~\eqref{eq:dd-for-X}, set this equal to $0$ and solve for some representation of $X(\hat{\rho})$ in terms of other known variables.
Note first that 
\begin{align*}
    \partial_{\hat{\rho}}X = -\frac{1}{\tdelta{c}}\hat{\rho}^{-1/\tdelta{c}-1} = -\frac{1}{\tdelta{c}}XX^{\tdelta{c}},
\end{align*}
and that we can re-write Eq.~\eqref{eq:dd-for-X} as a product of three functions of $X$:
\begin{align*}
    F_1 &\equiv NX^{\tdelta{c}}\\
    F_2 &\equiv [\eta X + \tdelta{c} X^2]\\
    F_3 &\equiv \bigexp{-A' X^2 + 2B'X - C'}.
\end{align*}
The derivative will thus be calculated structurally as 
\begin{align*}
    \partial_{\hat{\rho}}F_1 F_2 F_3 = F_1 F_2 \partial_{\hat{\rho}} F_3 + F_1 (\partial_{\hat{\rho}} F_2) F_3 + (\partial_{\hat{\rho}} F_1) F_2 F_3.
\end{align*}
Evaluating each of these derivatives in turn, we find
\begin{align*}
    \partial_{\hat{\rho}} F_1 &= N \partial_{\hat{\rho}} X^{\tdelta{c}} = N\tdelta{c} X^{\tdelta{c} -1}\partial_{\hat{\rho}} X = -NX^{\tdelta{c}-1}XX^{\tdelta{c}} = -NX^{2\tdelta{c}} \\ &= -F_1X^{\tdelta{c}},
\end{align*}
\begin{align*}
    \partial_{\hat{\rho}} F_2 &= \eta \partial_{\hat{\rho}} X + \tdelta{c} \partial_{\hat{\rho}} X^2\\
    &= -\frac{\eta}{\tdelta{c}}XX^{\tdelta{c}} - 2X(XX^{\tdelta{c}})\\
    &= -\frac{X^{\tdelta{c}}}{\tdelta{c}}(\eta X + 2\tdelta{c} X^2)\\
    &= -F_2 \cdot \left(\frac{X^{\tdelta{c}}}{\tdelta{c}}\frac{\eta X + 2\tdelta{c} X^2}{\eta X + \tdelta{c} X^2}\right), 
\end{align*}
and
\begin{align*}
        \partial_{\hat{\rho}} F_3 &= F_3 \partial_{\hat{\rho}}\left(-A' X^2 + 2B'X - C'\right)\\
        &= F_3\left(A'2X\frac{1}{\tdelta{c}}XX^{\tdelta{c}} - 2B'\frac{1}{\tdelta{c}}XX^{\tdelta{c}}\right)\\
        &= F_3 \frac{2X^{\tdelta{c}}}{\tdelta{c}}(A'X^2-B'X).
\end{align*}
Now, $\partial_{\hat{\rho}}F_1 F_2 F_3$ will evaluate to zero only if
\begin{align*}
    \nonumber &\frac{2X^{\tdelta{c}}}{\tdelta{c}}(A'X^2-B'X) -\left(\frac{X^{\tdelta{c}}}{\tdelta{c}}\frac{\eta X + 2\tdelta{c} X^2}{\eta X + \tdelta{c} X^2}\right) - X^{\tdelta{c}} = 0.
\end{align*}
This implies
\begin{align}
    \nonumber \ & 2(A'X^2-B'X) -\left(\frac{\eta X + 2\tdelta{c} X^2}{\eta X + \tdelta{c} X^2}\right) - \tdelta{c} = 0\\
    \nonumber \Longrightarrow\ & 2(A'X^2-B'X)(\eta X + \tdelta{c} X^2) - (\eta X + 2\tdelta{c} X^2) \\&\nonumber- \tdelta{c} (\eta X + \tdelta{c} X^2) = 0\\
    \nonumber \Longrightarrow\ & 2A'\tdelta{c} X^3 + (2A'\eta - 2B'\tdelta{c})X^2 - (2B'\eta + 2\tdelta{c} + \tdelta{c}^2)X \\&- \eta(1+\tdelta{c}) = 0,
\end{align}
a cubic function whose solutions represent the most probable profile of the properly transformed double distribution, $\hat{\rho}_{tf}$, where we may use whichever form of $S(m)$ we like in arriving at the constants $A'$ and $B'$.
\end{appendix}

\end{document}